\documentclass[journal,comsoc]{IEEEtran}

\usepackage{cite}
\usepackage{graphics}
\usepackage{epsfig}
\usepackage{url}
\usepackage{enumerate}
\usepackage{booktabs}
\usepackage{multirow}
\usepackage{bm}
\usepackage{bbm}
\usepackage{commath}
\usepackage{amsfonts,amsthm,amssymb,amsmath}
\usepackage{algorithm}
\usepackage{algorithmic}
\usepackage{float}
\usepackage{comment}

\newtheorem{theorem}{\textbf{Theorem}}

\graphicspath{{./Figures}}
\usepackage{subcaption}

\begin{document}
\bibliographystyle{IEEEtran}

\title{Achieving Scalable Capacity in Wireless Mesh Networks}

\author{Lei Lei, Aimin Tang and Xudong Wang~\IEEEmembership{Fellow,~IEEE}\\
\thanks{The authors are with the University of Michigan–Shanghai Jiao Tong University Joint Institute, Shanghai Jiao Tong University, Shanghai 200240 China. Corresponding author: Xudong Wang, e-mail: wxudong@ieee.org.}
}

\maketitle

\begin{abstract}
Wireless mesh networks play a critical role in enabling key networking scenarios in beyond-5G (B5G) and 6G networks, including integrated access and backhaul (IAB), multi-hop sidelinks, and V2X.
However, it still poses a challenge to deliver scalable per-node throughput via mesh networking. As shown in Gupta and Kumar's seminal research \cite{kumar}, multi-hop transmission results in a per-node throughput of $\Theta(1/\sqrt{n\log n})$ in a wireless network with $n$ nodes, significantly limiting the potential of large-scale deployment of wireless mesh networks. Follow-up research has achieved $O(1)$ per-node throughput in a dense network, but how to achieve scalability remains an unresolved issue for an extended wireless network where the network size increases with a constant node density. This issue prevents a wireless mesh network from large-scale deployment. To this end, this paper aims to develop a theoretical approach to achieving scalable per-node throughput in wireless mesh networks.
First, the key factors that limit the per-node throughput of wireless mesh networks are analyzed, through which two major ones are identified, i.e., link sharing and interference. Next, a multi-tier hierarchical architecture is proposed to overcome the link-sharing issue. 
The inter-tier interference under this architecture is then mitigated by utilizing orthogonal frequency allocation between adjacent tiers, while the intra-tier interference is reduced by considering two specific transmission schemes, one is MIMO spatial multiplexing with time-division, the other is MIMO beamforming.
Theoretical analysis shows that, the multi-tier mesh networking architecture can achieve a per-node throughput of $\Theta(1)$ in both schemes, as long as certain conditions on network parameters including bandwidth, antenna numbers, and node numbers of each tier are satisfied.
A case study on a realistic deployment of 10,000 nodes is then carried out, which demonstrates that a scalable throughput of $\Theta(1)$ is achievable with a reasonable assumption on bandwidth and antenna numbers.
\end{abstract}


\section{Introduction}
\IEEEPARstart{W}{ireless} mesh networking has recently emerged as a key technology in many wireless communication systems, where data is transmitted from the source to the destination in a multi-hop way, offering several prominent advantages such as flexibility, cost efficiency, and low complexity \cite{Wang_Survey}. One potential application of wireless mesh networking is to support backhauling of 6G networks and provide a more flexible service \cite{mesh_6G}. Furthermore, mesh networking is commonly used in vehicle-to-vehicle (V2V), vehicle-to-infrastructure (V2I), and vehicle-to-everything (V2X) networks as it can improve signal propagation and extend network coverage \cite{mesh_V2V, mesh_V2I}. It will be an enabling technology for multi-hop sidelinks~\cite{3gppTR_38.836}.

A primary concern in wireless mesh networking is how to achieve scalable end-to-end throughput in large-scale deployments. Pioneering work by P. Gupta and P. R. Kumar \cite{kumar} established a theory to analyze the capacity scaling law of wireless networks. Their results reveal that in a wireless network with $n$ nodes independently and uniformly distributed within a unit disk, each node can attain a per-node throughput of $\Theta(1/\sqrt{n\log{n}})$ via a multi-hop transmission strategy. However, even under optimal conditions, the per-node throughput cannot exceed $\Theta(1/\sqrt{n})$, leading to a decline in the per-node throughput of wireless networks by at least an order of $1/\sqrt{n}$ when using multi-hop transmission. This decline indicates that the throughput of multi-hop networks is not scalable as $n$ increases. In \cite{gap}, the $1/\sqrt{\log n}$ factor in the achievable per-node throughput is removed with the aid of paths that percolate across the network, i.e., a throughput of $O(1/\sqrt{n})$ is achievable.

In a dense network where the network size maintains unchanged while node density keeps increasing, the degradation in performance using multi-hop transmission is mainly caused by the interference due to concurrent wireless transmissions, as reported in \cite{kumar}.
Nevertheless, by employing a hierarchical cooperation (HC) scheme \cite{hc} that facilitates joint transmission and reception among nodes or exploiting node mobility in a wireless mobile network \cite{mobility}, a scalable throughput of $O(1)$ can be achieved. A more general conclusion is drawn in \cite{tse}, where four different operating regimes are identified based on the node density and path loss exponent. Besides, each of these regimes corresponds to a specific order-optimal transmission scheme.

In an extended network where the network size scales linearly with the number of nodes while maintaining a constant node density, the scaling law differs. Typically, in such networks, each node is capable of transmitting data only to its neighboring nodes, thereby generating negligible interference to far-off receiving nodes. 
It is revealed in \cite{tse} that multi-hop transmission is an optimal scheme in the extended network, which typically operates in the power-limited region and can achieve a  throughput of $O(1/\sqrt{n})$.
In \cite{xie}, a similar theoretic bound of $\Theta(1/\sqrt{n})$ is established for extended networks with nodes location satisfying a minimum distance constraint and a power-law path loss model with exponent $\alpha > 6$, or an exponential attenuation.
An upper bound of $O(1/n^{1/2-1/\alpha})$ is derived in \cite{OL_Upper_Bound}.
In general, a scalable throughput of $O(1)$ is not achievable in an extended network.


Consequently, how to achieve scalable per-node throughput remains an open issue for wireless mesh networks.
This paper aims to develop a method to resolve this issue. To this end, the throughput of an extended mesh network is first analyzed by taking a similar approach of existing work, but from the perspective of identifying the key factors that restrict the scalability of extended mesh networks.
The analysis reveals two main factors that limit scalability: link-sharing and interference. 
While the interference caused by concurrent nearby transmissions is a well-known issue in wireless networks, the link-sharing issue arises in a wireless mesh network because one node needs to help relay multiple data flows of other source-destination (S-D) pairs.
As $n$ increases, the number of data flows one node needs to relay also increases, which reduces the per-node throughput and limits the scalability.

To resolve the link-sharing issue of extended networks, the key approach is to reduce the number of data flows each node needs to relay.
A simple way is to enlarge the transmission range of all wireless nodes to be comparable with the network radius, i.e., the transmit power of each node should also grow as $n$ increases.
Besides, a sophisticated design of the interference management scheme is required to reduce excessive interference.
Hence, the above method is not feasible for practical deployment.
To address this issue, a multi-tier hierarchical architecture is proposed for extended mesh networks in this paper. 
It includes multiple types of relay nodes, being organized into multiple tiers and supporting data transfer for regular mesh nodes.
Data packets of S-D pairs are routed in the multi-tier mesh network using a routing policy named $D$-hop maximum routing. 
To improve the capacity scaling laws of a multi-tier mesh network, multi-input multi-output (MIMO) technology is utilized for data transmission of each link.
In particular, two transmission schemes are considered to increase the per-node throughput.
One is the spatial multiplexing scheme where the DoF gain of MIMO is exploited to enhance link capacity, and the other is the beamforming scheme where the power gain of MIMO is utilized to increase the transmission range of wireless nodes and thus further alleviate the link-sharing issue.

To resolve the remaining issue of interference, orthogonal frequency bands are allocated to adjacent tiers to avoid inter-tier interference, while time division or beamforming is used to reduce intra-tier interference.
More specifically, under the spatial multiplexing scheme, time division is used to separate data transmissions of adjacent nodes to reduce the excessive interference of concurrent data transmissions, while under the beamforming scheme, MIMO beamforming technology is utilized to transmit data with thin beams, without causing too much interference to nearby nodes.

Theoretical analysis shows that, a scalable throughput of $\Theta(1)$ can be achieved in both spatial multiplexing and beamforming schemes, as long as the scaling orders of bandwidth, antenna number, and node number at each tier satisfy certain requirements.
A case study is carried out subsequently, which reveals that the aforementioned requirements on network parameters are reasonable and achievable in realistic deployment, by leveraging high-frequency communications 
to support both large bandwidth and multiple antennas.

The key contributions of this paper are summarized as follows:
\begin{itemize}
    \item Two critical factors that limit the scalability of wireless networks are revealed by deriving and analyzing the achievable per-node throughput of a single-tier mesh network.

    \item A multi-tier hierarchical network architecture is proposed to tackle the link-sharing issue for extended mesh networks.
    
    \item Various physical layer technologies are incorporated into the multi-tier network architecture, and then the achievable per-node throughputs in different cases are derived.
        
    \item Theoretical analysis shows that scalable per-node throughput can be achieved via the multi-tier mesh network architecture along with certain physical layer technologies.
\end{itemize}

The rest of this paper is organized as follows.
A single-tier hexagonal mesh network is analyzed in Section \ref{sec:single-tier_throughput}, where two main factors that limit the scalability of mesh networks are identified.
To resolve the scalability issue, a multi-tier hierarchical architecture for mesh network is proposed in Section \ref{sec:multi-tier}.
The achievable per-node throughput for a multi-tier mesh network considering MIMO technologies is derived in Section \ref{sec:multi-tier_throughput}. In Section \ref{sec:scalability}, the conditions to achieve scalability are discussed.
A case study on realistic deployment is carried out to demonstrate the feasibility of the multi-tier mesh architecture.
Finally, the paper is concluded in section \ref{sec:conclusion}.

\section{Factors that Limit the Scalability of Single-Tier Mesh Networks}
\label{sec:single-tier_throughput}
Deriving scaling laws for wireless mesh networks has been the subject of extensive research.
However, most existing works fail to explicitly explain why wireless mesh networks suffer from scalability issues.
Addressing this issue is the primary objective of this section. 
By deriving and analyzing the scaling laws for the per-node throughput of wireless mesh networks, interference and link-sharing are identified as the two key factors that limit the scalability of wireless mesh networks.

\subsection{Network Model}
\begin{figure}[t]
	\centering
	\includegraphics[width=0.8\linewidth]{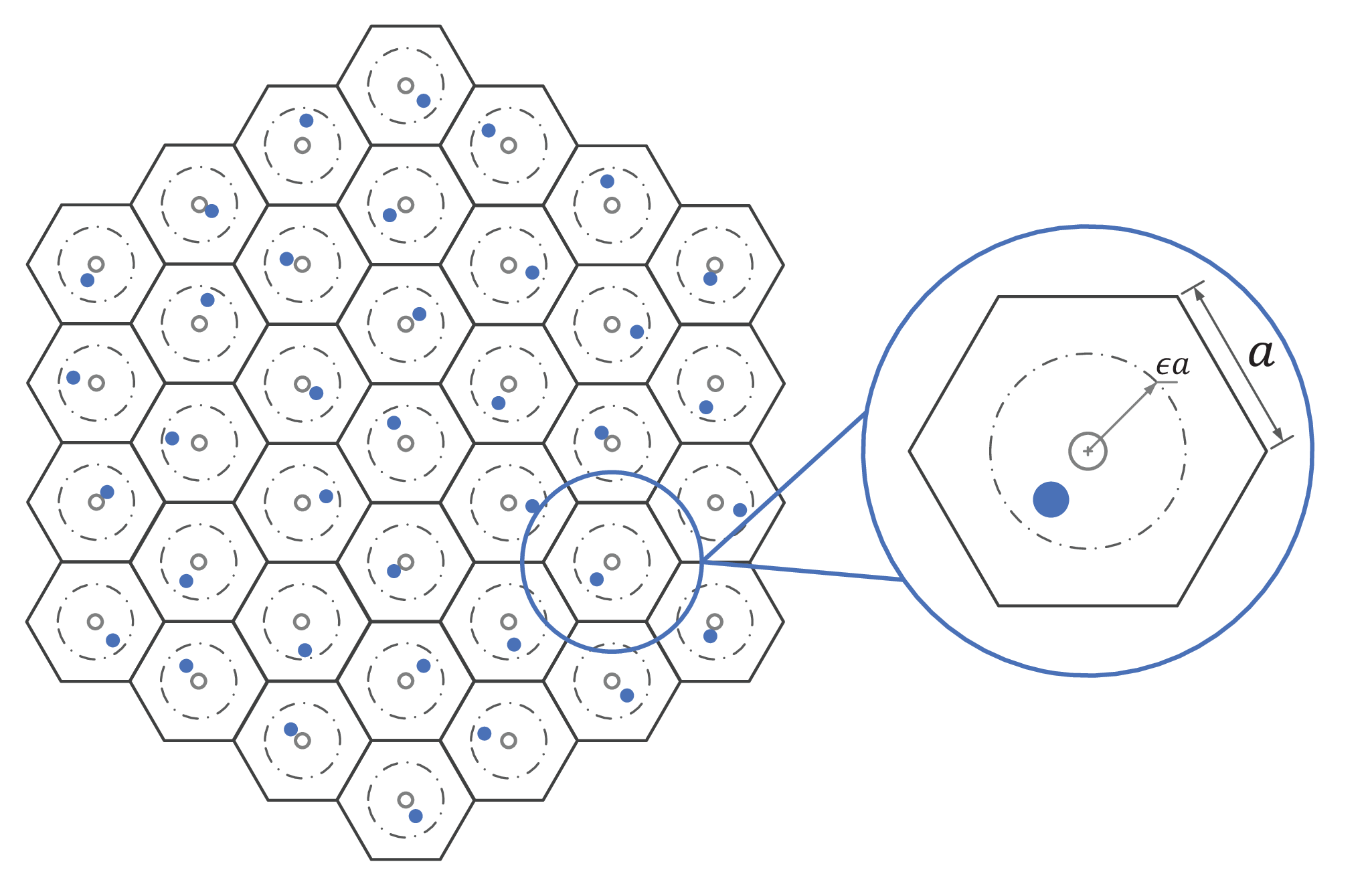}
    \caption{Hexagonal mesh network model.}
    \label{fig:network}
    \vspace{-12pt}
\end{figure}

Consider a wireless mesh network comprised of $n$ mesh nodes, as illustrated in Fig.~\ref{fig:network}. 
The network is divided into $n$ hexagonal cells, where each hexagonal cell contains exactly one mesh node.
Each mesh node can be placed either exactly in the center of the hexagonal cell (i.e., regular mesh network) or randomly distributed around the center within a small perturbation range (i.e., randomly perturbed network). 
Notably, different from the random network in \cite{kumar} where cell size has to be at least $\log n$ to make sure that each cell contains at least one node with high probability (w.h.p.), the geographical locations considered in this paper is more like a regular lattice. 
Consequently, the $1/\sqrt{\log n}$ factor that occurred in the achievable per-node throughput in \cite{kumar} can be removed, thus eliminating the effect of nodes' locations on the scalability.
Define the $l$-th outer ``ring'' of a cell as the set containing cells that are $l$ hops away from the given cell. 
For the central cell, there are $6l$ cells in the $l$-th outer ring. 
Suppose there are $L_{r}$ outer rings around the central cell in total.
The total number of nodes is then $n=3L_{r}\left(L_{r}+1\right)+1$.
The distance between the centers of two neighboring cells is $\sqrt{3}a$, where $a$ denotes the side-length of the hexagonal cell, and the radius of the mesh network can be expressed as $\sqrt{3}L_{r}a$. 
For the extended single-tier mesh network, when $n$ increases, the cell size remains unchanged, i.e., $a$ is a fixed constant, while the network size grows linearly with $n$.
Thus, it can be derived that $L_r = \Theta(\sqrt{n})$.

In terms of data traffic, each node randomly and independently selects another node as its destination, forming a source-destination (S-D) pair. 
The number of S-D pairs $N_{S-D}$ is equal to $n$, i.e., $N_{S-D} = n$. 
There are $N_{S-D}$ data flows across the mesh network in total.

\subsection{Transmission Model}
Let $d_{ij}$ denote the distance between nodes $i$ and $j$, and assume equal transmit power $P$ for all mesh nodes. The received signal power $P_{ij}$ at node $j$ from node $i$ can be expressed as 
\begin{equation}
    \label{equ:P_ij}
    P_{ij}=C P d_{ij}^{-\alpha},
\end{equation}
where $\alpha$ is the path loss exponent with typical values in outdoor environments of $2\leq \alpha \leq 4$, and $C$ is the effective antenna gain. 
To obtain a feasible transmission rate, the received signal power should be no less than a certain threshold, denoted by $P^{0}$, i.e., $P_{i,j} C P d_{ij}^{-\alpha} \ge P^{0}$, or equivalently,
\begin{equation}
    \label{equ:d_ij}
    d_{ij} \le \left( \frac{CP}{P^{0}} \right)^{1/\alpha},
\end{equation}
which provides a threshold on the maximum allowed distance for two mesh nodes to communicate.

In mesh networks, data transmissions are accomplished in multiple hops.
We assume that all mesh nodes have the same transmission range $r_0$. 
As the transmission range increases, the number of hops required decreases. 
Let $d_{l,max}$ and $d_{l,min}$ be the maximum and minimum distances from a node to another node located in the $l$-th outer ring around itself, whose concrete values can be found in \cite{Lei_MASS2022}. 
A node can communicate with another node inside the $r$-th outer ring around itself if
\begin{equation*}
    \label{equ:r0_k}
    d_{r,max} \le r_0 < d_{r+1, min}, \quad r = 1,...,L_{r}.
\end{equation*}
This condition for transmission range can be equivalently transformed into requirements on transmit power.
Here we only consider the minimum required power to provide a feasible communication rate, i.e., $CP d_{r,max}^{-\alpha} = P^{0}$.
Since the network radius $a$ remains constant in the extended network, $d_{r,max}$ grows in the same order with $r$.
Thus, the required transmit power can be derived as
\begin{equation}
    \label{equ:P_single}
    P = P^{0} d_{r,max}^{\alpha} / C = \Theta \left( r^{\alpha} \right).
\end{equation}

Denote by $R_{ij}$ the transmission rate from node $i$ to node $j$.
$R_{ij}$ can be obtained by Shannon's formula, i.e., 
\begin{equation}
    \label{equ:R_ij}
    R_{ij} = W \log_2\left(1+\gamma_{ij}\right),
\end{equation}
where $W$ is the allocated channel bandwidth, $\gamma_{ij}$ represents the Signal-to-Interference-plus-Noise Ratio (SINR) of the received signal at node $j$ from node $i$.

\subsection{Throughput Analysis of Single-Tier Mesh Networks}
In this paper, we adopt the same definition of per-node throughput as \cite{kumar}.
An achievable per-node throughput of a single-tier mesh network, denoted by $R(n)$, indicates that each node can transmit data to its chosen destination at a rate of $R(n)$. 
To address the factors that limit the scalability of mesh networks, an investigation into the scaling law of per-node throughput for single-tier mesh networks, i.e., how $R\left(n\right)$ changes with $n$, is essential. 
In particular, two cases characterizing two different transmission range $r$ are analyzed in this paper to examine the per-node throughput of single-tier mesh networks, namely $r=\Theta \left( 1 \right)$ and $r=\Theta \left( \sqrt{n} \right)$. 
In the former case, nodes can only access other nodes in their neighboring cells, and multi-hop transmission is required for data communication with distant nodes. 
On the other hand, in the latter case, data transmission can be completed within several hops, since the transmission range is in the same order as the network radius. 
Thus, we term these two specific transmission schemes as Short-Hop (SH) scheme and Long-Hop (LH) scheme, respectively.
According to (\ref{equ:P_single}), the required transmit power under these two schemes should scale as
\begin{equation}
    \label{equ:P_single_scaling}
    P = \begin{cases}
        \Theta (1), & \text{for SH scheme}, \\
        \Theta \left( n^{\alpha/2} \right), & \text{for LH scheme}.
    \end{cases}
\end{equation}
The per-node throughputs of the SH and LH transmission schemes are stated in the following theorem.

\begin{theorem} (Single-Tier)
\label{thm:R(n)_Single_Tier}
For both regular and randomly perturbed hexagonal mesh networks with $n$ nodes, the achievable per-node throughput is
\begin{equation*}
    R\left(n\right) 
    = \begin{cases}
        \Theta\left(W/\sqrt{n}\right), & r=\Theta \left(1\right), \\
        \Theta\left(W/n\right), & r=\Theta \left(\sqrt{n}\right), \\
    \end{cases}
\end{equation*}
where $W$ is the bandwidth allocated to wireless transmission, $r=\Theta \left(1\right)$ and $r=\Theta \left( \sqrt{n} \right)$ correspond to SH and LH transmission schemes, respectively.
\end{theorem}

In general, the procedure of obtaining per-node throughput of single-tier mesh networks can be partitioned into the following steps. 
First, a lower bound on the transmission rate between two communicating mesh nodes, denoted by $R^{(L)}$, is derived.
Next, the maximum number of interfering neighboring cells, denoted by $\Delta c$, is obtained.
Utilizing time division to separate these interfering transmissions, each cell is active in every $\left(1+\Delta c\right)$ time slots \cite{kumar}.
An upper bound on the number of data flows a node needs to relay is then derived as $Z^{(U)}$.
Last but not least, assuming these no more than $Z^{(U)}$ data flows are also separated using time division, an achievable per-node throughput can be obtained as
\begin{equation}
    \label{equ:R(n)_ini_single}
    R\left(n\right)=\frac{R^{(L)}}{\left( 1+\Delta c \right)Z^{(U)}}.
\end{equation}
In the sequel, the scaling laws on the per-node throughput of single-tier mesh networks will be derived in detail.

\subsubsection{Lower bound on transmission rate}
To derive a lower bound on the transmission rate, we need to derive the minimum received power and maximum interference.
As for the data transmission from node $i$ to node $j$, the cumulative interference suffered by node $j$ comes from all the other concurrent transmissions sharing the same frequency bandwidth. 
The maximum interference occurs when the node $j$ is located at the center cell of the network and all other nodes transmit concurrently.
Using the time division method, the SINR of data transmission is $\Omega(1)$ in a single-tier mesh network, as shown in \cite{ish} and \cite{Lei_MASS2022}.
Thus, the transmission rate between node $i$ and node $j$ is $R_{ij}=W \log_2(1+\gamma_{ij}) = \Omega(W)$.
A lower bound on transmission rate between two communicating nodes for the single-tier mesh network can be obtained as
\begin{equation}
    \label{equ:R_L_single}
    R^{(L)} = \Theta(W).
\end{equation}

\subsubsection{Number of Interfering Cells}
Recall the network model shown in Fig.~\ref{fig:network}. 
Since each node can communicate with another node inside the $r$-th outer ring around itself, in the worst case, all other nodes inside the $r$-th outer ring around the receiving node except the transmitting node are considered to be interfering nodes.
The number of interfering cells for a receiving node scales in the order of $r^2$, and hence
\begin{equation}
    \label{equ:Delta_c_single}
    \Delta c = \Theta\left(r^2\right) = 
    \begin{cases}
        \Theta\left(1\right), & r=\Theta \left(1\right), \\
        \Theta\left(n\right), & r=\Theta \left(\sqrt{n}\right). \\
    \end{cases}
\end{equation}

\subsubsection{Number of Data Flows Each Node Relay}

The number of hops needed for data transmission depends on transmission range $r_0$.
A larger transmission range leads to fewer hops needed for data transmission.
Let $Z_i$ denote the number of data flows node $i$ participates, it has been derived in \cite{Lei_MASS2022} that
\begin{equation*}
    \mathbb{E}\left[Z_i\right]=
    \begin{cases}
        \Theta\left(\sqrt{n}\right), & r=\Theta \left(1\right), \\
        \Theta\left(1\right), & r=\Theta \left(\sqrt{n}\right). \\
    \end{cases}
\end{equation*}
Intuitively, when each node can only access another node in a neighboring cell, the expected number of hops needed for each data flow scales in the same order as the number of outer rings of the central cell, i.e., $\Theta(sqrt{n})$.
The total number of hops across the network is $\Theta(n\sqrt{n})$, and thus each node needs to relay $\Theta(\sqrt{n})$ data flows on average.

For SH transmission scheme where $r=\Theta \left(1\right)$, an upper bound on $Z_i$ can be obtained by applying Chernoff upper tail bound \cite{chernoff} to $\mathbb{E}\left[Z_i\right]$.
The result shows that $Z_i \le \left(1+\delta \right) \mathbb{E}\left[ Z_i \right]$ for every node $i$ with the probability no smaller than $1-1/n^2$, where $\delta=\sqrt{6\log n/\mathbb{E}\left[Z_i\right]}$. 
Thus, the upper bound on $Z^{(U)}$ under the SH transmission scheme can be obtained as
\begin{equation}
    \label{equ:Z_upper_o}
    Z^{(U,SH)} =\left(1+\delta \right) \mathbb{E}\left[ Z_i \right]=\Theta\left(\sqrt{n}\right).
\end{equation}
As for LH scheme where $r=\Theta \left(\sqrt{n}\right)$, since most data transmissions are finished within a few long-range hops, the number of data flows each node needs to relay is upper bounded by a constant, i.e.,
\begin{equation}
    \label{equ:Z_upper_Omega}
    Z^{(U,LH)} = \Theta\left(1\right).
\end{equation}
Combining (\ref{equ:Z_upper_o}) and (\ref{equ:Z_upper_Omega}) yields that
\begin{equation}
    \label{equ:Z_U_single}
    Z^{(U)}=
    \begin{cases}
        \Theta\left(\sqrt{n}\right), & r=\Theta \left(1\right), \\
        \Theta\left(1\right), & r=\Theta \left(\sqrt{n}\right), \\
    \end{cases}
\end{equation}

\subsubsection{Capacity scaling law}
Plugging (\ref{equ:R_L_single}), (\ref{equ:Delta_c_single}), and (\ref{equ:Z_U_single}) into (\ref{equ:R(n)_ini_single}), the achievable per-node throughput for the single-tier hexagonal mesh network can be obtained as
\begin{equation}
    \label{equ:R(n)_single}
    \begin{aligned}
        R\left(n\right) 
        &\ge \frac{R^{(L)}}{\left(1+\Delta c\right)Z^{(U)}} \\
        &= \begin{cases}
            \Theta\left(W/\sqrt{n}\right), & r=\Theta \left(1\right), \\
            \Theta\left(W/n\right), & r=\Theta \left(\sqrt{n}\right). \\
        \end{cases}
    \end{aligned}
\end{equation}
This completes the proof of Theorem \ref{thm:R(n)_Single_Tier}.

\subsection{Key Factors that Limit the Scalability}
As shown in (\ref{equ:R(n)_single}), the per-node throughput decreases in the order of $1/\sqrt{n}$ and $1/n$ under SH and LH schemes, respectively.
By analyzing the derivation procedure, the following two factors that limit the scalability are obtained.

\textbf{Link-Sharing:} 
From (\ref{equ:R(n)_single}), the per-node throughput of mesh networks is lower bounded by $1/\sqrt{n}$ in the SH scheme where $r=\Theta (1)$. 
In this scenario, each node can only access a few nearby nodes and most data flows are delivered hop by hop.
Thus, many nodes need to help relay data of other S-D pairs.
Since every data flow is expected to be finished within $\Theta(\sqrt{n})$ hops, the total number of disjoint hops needed is $\Theta(n\sqrt{n})$.
However, the single-tier mesh network can only support $\Theta(n)$ links, thus leading to the link-sharing issue, i.e., each node needs to help relay $\Theta(\sqrt{n})$ data flows of other S-D pairs.
On the other hand, in the LH scheme where $r=\Theta (\sqrt{n})$, data transmissions can be accomplished in several hops, and each node needs to relay $O(1)$ data flows.
Thus, the link-sharing issue is no longer a key factor that limits the per-node throughput.

\textbf{Interference:} 
Since time division is used to reduce excessive interference, the number of interfering cells shows a direct impact on the per-node throughput.
Specifically, in the SH scheme where $r=\Theta \left(\sqrt{n}\right)$, data transmissions take place between neighboring cells.
The number of interfering cells is $O(1)$, which has little effect on the throughput of the SH scheme.
On the contrary, in the LH scheme where $r=\Theta (\sqrt{n})$, there are ${\Delta c}^{(LH)}=\Theta(n)$ interfering cells which need to keep silent for successful long-range transmission, which leads to the $1/n$ factor in the scaling law of throughput in LH scheme.

It can be seen that there exists a trade-off between link-sharing and interference, which is dependent on the transmission range. 
As the transmission range increases, the number of data flows that each node needs to relay may decrease, i.e., the link-sharing issue is alleviated, but it comes at the cost of a larger number of interfering cells.

\section{Architecture Design for Mesh Networks}
\label{sec:multi-tier}
As discussed in the previous section, link-sharing is the primary factor that limits the scalability of wireless mesh networks using the short-hop transmission scheme, corresponding to the extended network.
As a result, reducing the number of data flows each node needs to relay is the key approach to improving per-node throughput. 
While increasing transmit power to enlarge the transmission range can decrease the number of hops needed for transmission of each S-D pair, the required power for each node grows exponentially according to (\ref{equ:P_single_scaling}).
Besides, a sophisticated interference cancellation scheme like hierarchical cooperation \cite{hc} is required to mitigate the excessive interference due to the increase of transmission range.
Both of the above two factors make it impractical to solely increase transmit power in the realistic deployment.
Thus, a new efficient architecture is required for wireless mesh networks to resolve both the link-sharing and interference issues.

\subsection{Insights to Architecture Design}

In the previous analysis of per-node throughput in single-tier mesh networks, it has been established that link-sharing is the main cause of the decline in the per-node throughput of extended networks where the multi-hop transmission scheme is employed. 
To achieve scalability in extended mesh networks, alleviating the link-sharing issue is essential. 
As shown before, the link-sharing issue arises when a node is required to relay data of other S-D pairs. 
When the SH transmission scheme is used, the number of data flows that each node needs to relay is in the order of $\sqrt{n}$.
On the other hand, when the long-hop (LH) transmission scheme is used, each node only needs to relay $O(1)$ data flows, but the number of interfering cells becomes $\Theta(n)$, which also limits the per-node throughput.
Moreover, it has been proven in \cite{xie} and \cite{tse} that, the multi-hop transmission scheme is orderly optimal in extended networks.
Thus, a new architecture is required to address both the link-sharing issue and interference to achieve scalability in extended networks.

To address the link-sharing issue in an extended mesh network, the key is to limit the number of data flows each node needs to relay.
To achieve this goal, additional relay nodes can be exploited to establish more wireless links and facilitate data transmission. 
In this way, a two-tier mesh network is constructed where the first tier contains all data nodes and the second tier is comprised of relay nodes.
To alleviate the link-sharing issue in the data tier, the following routing policy can be utilized.
Specifically, for a specific S-D pair, if the required hop counts using a pure multi-hop scheme in the first tier exceeds a certain predetermined threshold, then the data will be transmitted to the relay tier in a multi-hop manner.
After reaching the relay node near the destination, the data will be sent to the destination node using multiple hops.
The link-sharing issue in the data tier can be resolved since each node only needs to relay $O(1)$ data flows regardless of the number of data nodes.
To achieve this target, the density of relay nodes should maintain as a constant independent of $n$ such that each node can access a relay node within a certain number of hops.
The number of required relay nodes grows in the order of $\Theta(n)$.
The relay tier can be regarded as a separate mesh network.
The link-sharing issue in the relay tier now becomes the bottleneck of throughput if the transmission range of relay nodes is fixed independent of $n$.
One way of resolving this issue is keeping the number of relay nodes constant while increasing their transmission range and transmission rate in accordance with $n$, which can be achieved by increasing transmit power, allocating more bandwidth, and employing more antennas.
Nevertheless, as $n$ increases, the required transmit power will become out-of-range in this two-tier mesh network.

As the link-sharing issue will also manifest in the relay tier of the aforementioned two-tier networks, which curtails the scalability of this type of two-tier mesh network.
Thus, a multi-tier hierarchical architecture for mesh networks is developed in this paper.
As depicted in Fig.~\ref{fig:net_hier}, several types of relay nodes with varying transmission ranges are overlaid with the data tier. 
Nodes at higher tiers have larger transmission ranges and higher transmission rates than lower-tier nodes, which is achieved by allocating more communication resources including bandwidth, antenna numbers, and transmit power.
The specific system model, scaling relations of network parameters, and the routing policy are presented in detail in the next section.

\subsection{Architecture Design}
\subsubsection{Node Distribution}

\begin{figure}[t]
	\centering
	\includegraphics[width=0.8\linewidth]{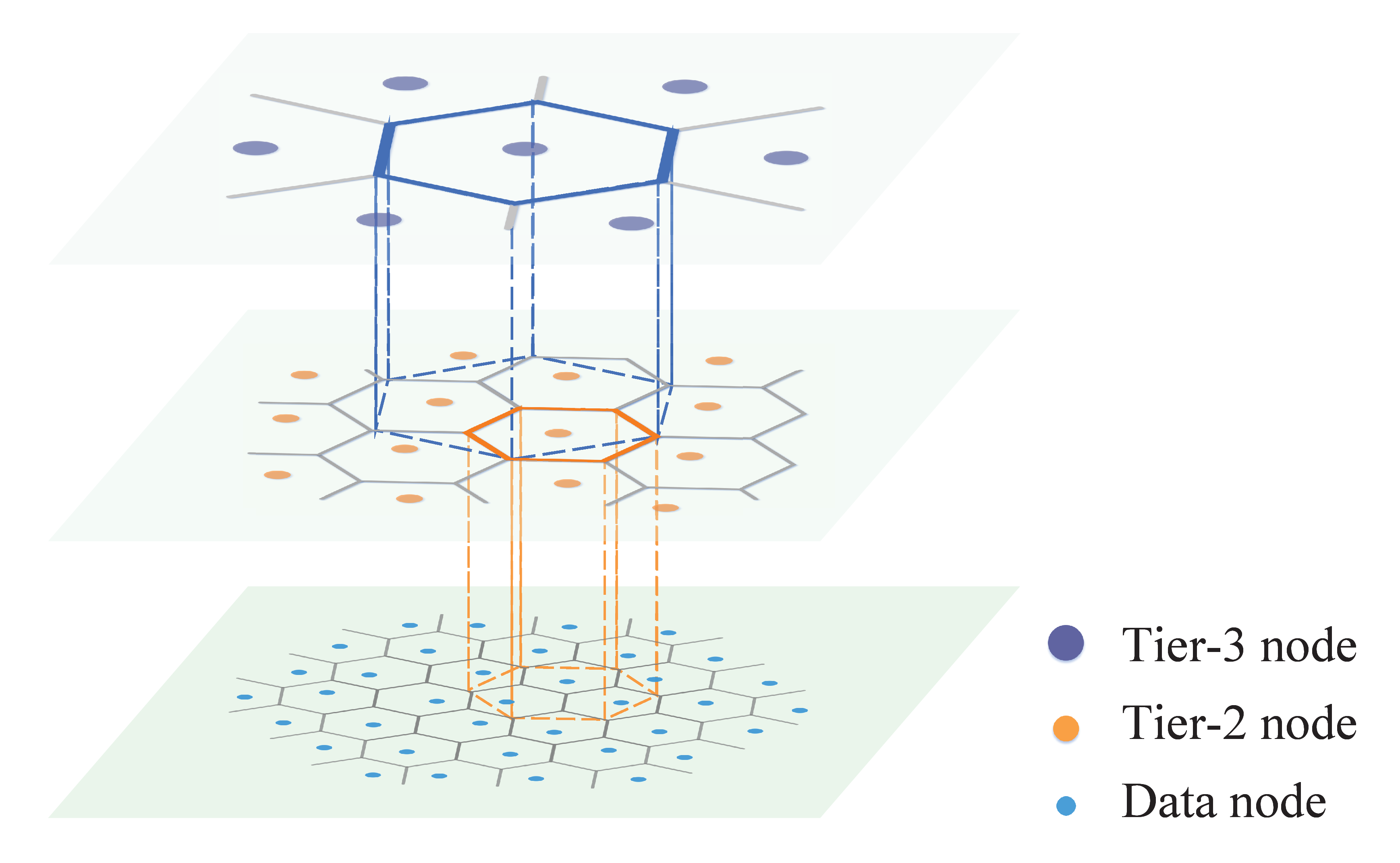}
    \caption{Illustration of the multi-tier hierarchical architecture.}
    \label{fig:net_hier}
    \vspace{-12pt}
\end{figure}

We consider a multi-tier wireless network, illustrated in Fig.~\ref{fig:net_hier}, where a hierarchical architecture is constructed with multiple relay tiers overlaid with the first tier which contains $n$ data nodes.
These relay nodes are deployed solely to aid in the transmission of data and do not generate any traffic data themselves.

Suppose there are $L$ tiers of nodes in total, with $n_l$ denoting the number of nodes in the $l$-th tier. It follows that $n_1 = n$.
At each tier, the network is divided into $n_l$ hexagonal cells, where each cell contains exactly one $l$-th tier node, i.e., the same setting as the aforementioned single-tier hexagonal mesh network.
Each node is located around the center of the cell with an allowed small random perturbation.
Communication of the $l$-th tier takes place over a bandwidth of $W_l$.
In addition, each node at the $l$-th tier is equipped with $M_l$ antennas.
The transmit power of each antenna element is denoted as $P_l$.

As for data traffic, it is still assumed that each data node in the first tier randomly and independently chooses another data node as its destination, i.e., an S-D pair.
Thus, there are still $N_{S-D}=n$ data flows.

\subsubsection{Scaling of Network Parameters}

\begin{table}[t]
    \centering
    \caption{Scaling Relation of Network Parameters}
    \label{tab:params_scaling}
    \begin{tabular}{c|c|c}
        \hline
        & Scaling & Parameter range  \\
        \hline
        Node number & $n_l = n / l^{k}$ & $k \in [2, \infty]$  \\
        Bandwidth & $W_l = W_1 l^{\psi}$ & $\psi \in [1, \infty]$ \\
        Antenna number & $M_l = M_1 l^{\upsilon}$ & $\upsilon \in [1, \infty]$  \\
        \hline
    \end{tabular}
    
    \vspace{-12pt}
\end{table}

In this paper, we consider the scaling relations for network parameters in terms of node number $n_l$, allocated bandwidth $W_l$, and antenna number $M_l$ with respect to the tier index $l$, as illustrated in Table \ref{tab:params_scaling}.
To ensure that the total number of nodes in the network is finite, $n_l$ should at least decrease in the square order of $l$.
For general investigation, we assume $n_l = n/l^{k}$ where $k \ge 2$.
Consequently, the cell size of the $l$-th tier, denoted by $A_l$, is $A_l = n / n_l = l^{k}$.
Ten, the cell radius of the $l$-th tier, denoted by $a_l$, is 
\begin{equation}
    \label{equ:a_l}
    a_l = \Theta \left( A_l \right) = \Theta \left( l^{k/2} \right).
\end{equation}
To avoid the link-sharing issue in the highest tier, data transmission is supposed to be finished within a few hops.
Thus, it should be satisfied that $n_L = n / k^{L} = \Theta(1)$, and the number of total tiers $L$ can be obtained as
\begin{equation}
    \label{equ:L}
    L = \Theta \left( \sqrt[k]{n}\right),
\end{equation}
which is dependent both on $n$ and $k$.

As explained before, the allocated bandwidth and antenna number should also increase as $n$ grows to maintain a non-decreasing throughput.
Here, we also assume bandwidth and antenna number are increasing in the polynomial order with respect to $l$, i.e., $W_l = l^{\psi}$ and $M_l = l^{\upsilon}$ where $\psi$ and $M_l$ are both constants no smaller than 1.
Thus, the total required bandwidth is
\begin{equation*}
    W_{\text{tot}} = \sum_{l=1}^{L} W_l = \Theta \left( W_1 L^{\psi+1}\right) = \Theta \left( n^{(\psi+1)/k} \right).
\end{equation*}
The maximum required antenna number is
\begin{equation*}
    M_{\text{max}} = W_L = \Theta \left( M_1 L^{\upsilon} \right) = \Theta \left( n^{\upsilon /k} \right).
\end{equation*}
As can be seen, the required total bandwidth and maximum antenna number grow to infinity as $n$ increases.
Therefore, to obtain a reasonable and achievable setting for network parameters, the values of the three scaling orders, $\psi$, $\upsilon$, and $k$ need to be carefully chosen in realistic deployment, which will be later discussed in Section \ref{sec:scalability}.

Different from other parameters, $P_l$ should be chosen to maintain a feasible link rate between two communicating nodes, as in the case of single-tier mesh networks discussed in Section \ref{sec:single-tier_throughput}.
Based on (\ref{equ:P_single}), to guarantee a feasible link rate between two neighboring nodes in the $l$-th tier, the received signal should be no less than a certain threshold, denoted by $P_l^{0}$, i.e., $C_l P_l d_{l}^{-\alpha_l} \ge P_l^{0}$,
or equivalently, 
\begin{equation}
    \label{equ:P_l_con_ini}
    P_l \ge P_l^{0} d_{l}^{\alpha_l} / C_l,
\end{equation}
where $d_{l}$ represents the distance between two neighboring nodes in the $l$-th tier, $\alpha_l$ and $C_l$ are the path-loss exponent and the constant determined by frequency, antenna profile, etc. of the $l$-th tier, respectively.
From (\ref{equ:a_l}), we obtain that $d_{l} = \Theta( \sqrt{n / n_l} ) = \Theta( l^{k/2})$, and hence
\begin{equation}
    \label{equ:ratio_d}
    d_{l+1}/d_{l} = \left( 1+1/l \right)^{k/2}.
\end{equation}
Subscribing $d_{L}=\Theta(L^{k/2})=\Theta(\sqrt{n})$ into \eqref{equ:P_l_con_ini}, the scaling law for required transmit power of the highest relay nodes is
\begin{equation}
    \label{equ:P_L_con}
    P_L = \Omega \left( n^{\alpha_{L} 2} \right).
\end{equation}

\subsubsection{$D$-hop Maximum Routing Policy}

\begin{figure}[t]
	\centering
	\includegraphics[width=0.75\linewidth]{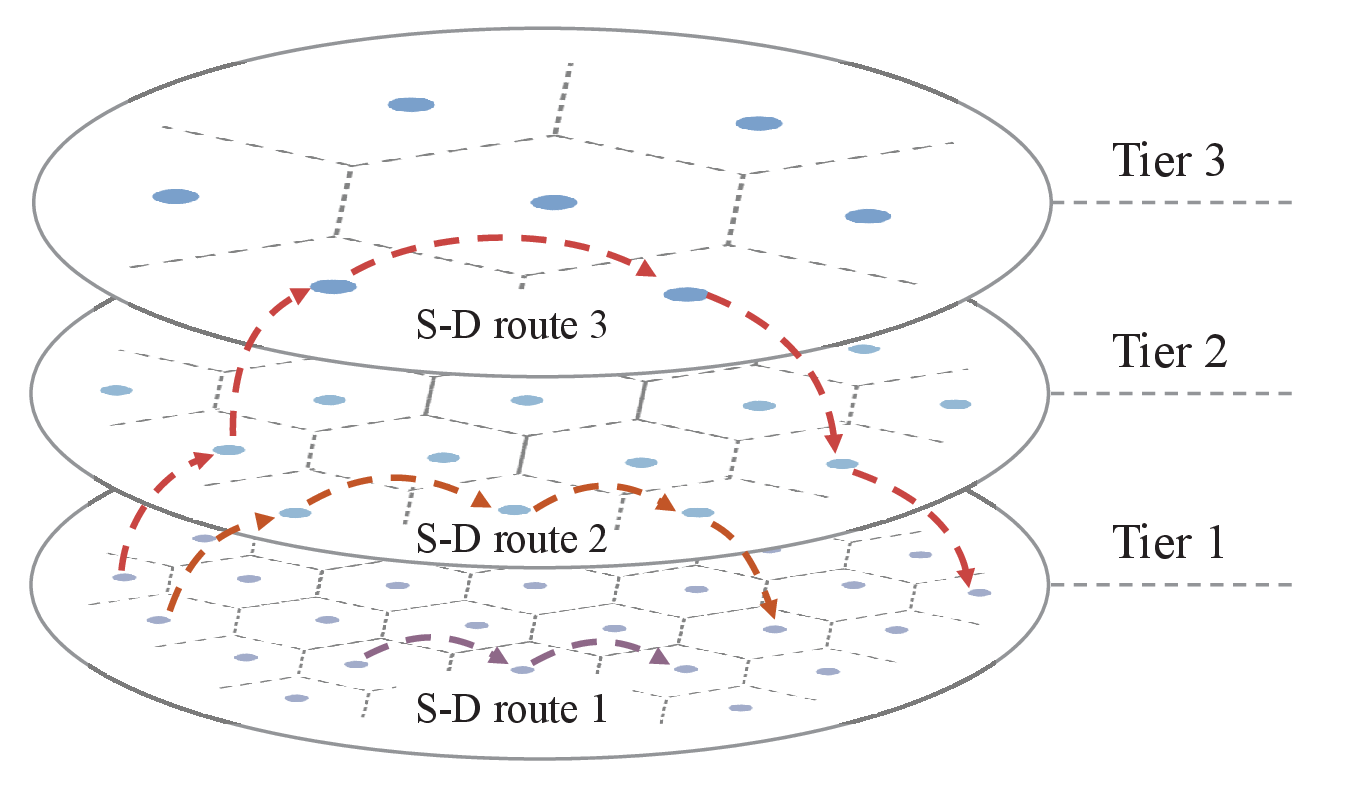}
    \caption{$D$-hop maximum routing policy.}
    \label{fig:tran_scheme}
    \vspace{-12pt}
\end{figure}


As previously noted, the crux of resolving the link-sharing issue entails a reduction in the number of data flows that each node assists in relaying. 
To leverage the transport capacity furnished by newly added multiple types of relay nodes, a tier-by-tier routing strategy is adopted to convey data from the source to the destination. 
The overarching principle of this routing policy is to effectuate data transmission for a given S-D pair in a tier-wise fashion across the multi-tier mesh network.

Considering each tier as an individual mesh network, it is feasible for each node to access its neighboring nodes within a single hop, allowing for data communication between nodes using multi-hop transmission at each tier. 
To prevent the link-sharing issue observed in the single-tier mesh network, it is crucial to restrict the number of hops for data transmission at each tier. 
Therefore, we propose a $D$-hop maximum routing policy for the multi-tier mesh network based on the $L$-maximum routing policy discussed in \cite{D-T_L-Hop, INFOCOM2011}. 
At each tier, the source node sends data to its corresponding destination node using the multi-hop scheme if it can be reached within $D_l$ hops. 
The specific values for $D_l$ will be obtained later.
Otherwise, the source node sends the data to its nearest next higher-tier node using the multi-hop transmission scheme. 
A similar decision is made for each subsequent tier until the data reaches the specific tier where data transmission can be finished within certain hops.
Next, the data will be transmitted to the first tier in a similar manner and finally arrive at the destination mesh node. The corresponding destination node for the above ``new'' source node at each relay tier is chosen as the same-tier relay node nearest to the next-lower tier destination node. 
Fig.~\ref{fig:tran_scheme} illustrates an example of how to achieve data transmission for an S-D pair under this tier-by-tier routing policy.

\begin{figure}[t]
	\centering
	\includegraphics[width=0.8\linewidth]{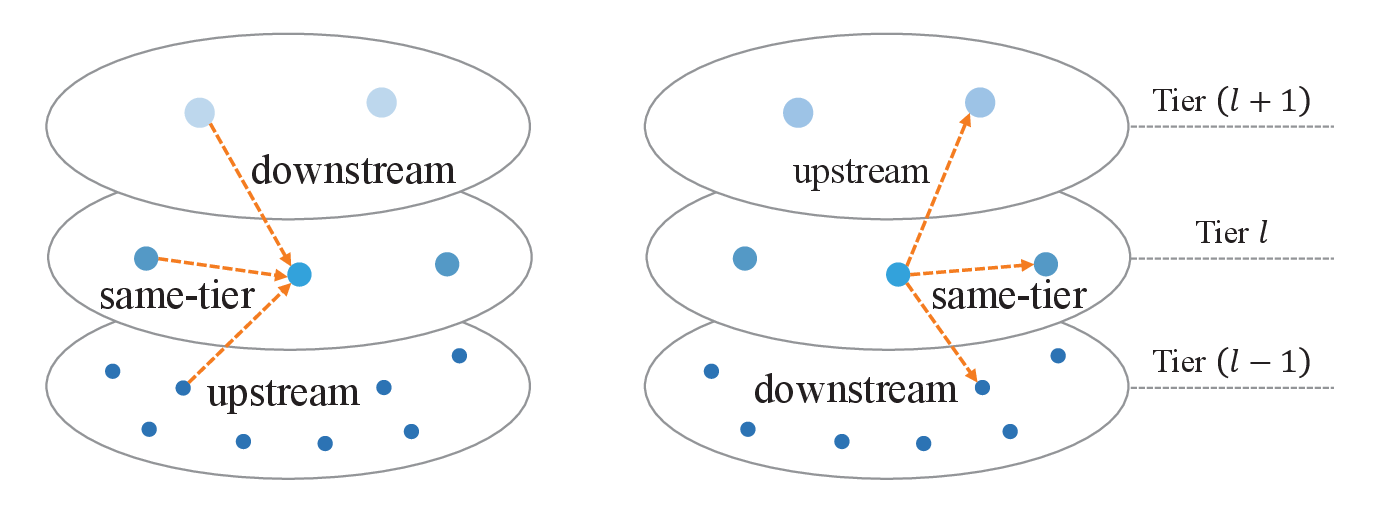}
    \caption{Data transmission and reception for a node at tier $l$.}
    \label{fig:tran_recv_l}
    \vspace{-12pt}
\end{figure}

In addition, Fig.~\ref{fig:tran_recv_l} depicts the three types of data transmission and reception for a node in the $l$-th tier ($2 \le l \le L-1$).    
Data transmission of the $l$-th tier comprises same-tier transmission, upstream transmission to the $(l+1)$-th tier, and downstream transmission from the $(l+1)$-th tier. 
These transmissions consume the same bandwidth allocated to the $l$-th tier.
As for the $L$-th tier, data transmission contains only the same-tier transmission.

%

\section{Achievable Per-Node Throughput for A Multi-tier Mesh Network}
\label{sec:multi-tier_throughput}
In this section, the achievable per-node throughput for the proposed multi-tier mesh network is established under the $D$-hop maximum routing policy.
To be specific, section \ref{subsec:throughput_def} extends the definition of per-node throughput to the case of the multi-tier mesh network.
Section \ref{subsec:transmission_scheme} briefly introduces the two transmission schemes for the multi-tier mesh network studied in this paper.
Specifically, based on the usage of multiple antennas, two transmission schemes, i.e., spatial multiplexing and beamforming schemes, are considered.
Finally, the per-node throughput under both transmission schemes is derived in section \ref{subsec:multi-tier_throughput}.

\subsection{Per-Node Throughput of A Multi-Tier Mesh Network}
\label{subsec:throughput_def}

We first extend the definition of per-node throughput of a single-tier mesh network to the case of a multi-tier mesh network.
Similar to the single-tier mesh network, a per-node throughput of $R(n)$ bps, for a multi-tier mesh network of $n$ wireless data nodes, is said to be achievable if each node can transmit data to its chosen destination at a rate of $R(n)$ bps.
Moreover, an end-to-end rate of $R_l(n)$ bits per second for the $l$-th tier of the multi-tier mesh network, is defined to be achievable if any two nodes in the $l$-th tier can communicate with each other at a rate of $R_l(n)$ bps.

The procedure of deriving an achievable per-node throughput in the multi-tier mesh network is summarized in the following.
Let $R_l^{(L)}$ denote the lower bound on the transmission rate of the $l$-th tier's links, including the same-tier transmission, the upstream transmission to the $(l+1)$-th tier, and the downstream transmission from the $(l+1)$-th tier.
Let $\Delta c_l$ and $Z_l^{(U)}$ denote the maximum number of interfering neighbors for a node in the $l$-th tier and the maximum number of data flows each node in the $l$-th tier needs to relay, respectively.
Since time division is utilized to reduce excessive intra-tier interference and separate data transfers of multiple data flows at one node, the achievable end-to-end rate between two nodes at the $l$-th tier, denoted by $R_l(n)$, can be obtained as
\begin{equation}
    \label{equ:R_l(n)_ini}
    R_l(n) \ge \frac{R_l^{(L)}}{(1+\Delta c_l)Z_l^{(U)}}.
\end{equation}
Under the $D$-hop maximum routing policy, every data flow is finished in a tier-by-tier manner, given the achievable per-node throughput of the $l$-th tier $R_l^{(L)}$ in the multi-tier mesh network, the achievable end-to-end rate for every S-D data flow, or per-node throughput for the multi-tier mesh network $R(n)$, can be obtained by taking minimum on all $R_l^{(L)}$ from $l=1$ to $L$, i.e., 
\begin{equation}
    \label{equ:R(n)_def}
    R(n) = \min_{1 \le l \le L} R_l(n),
\end{equation} 

\subsection{Transmission Scheme}
\label{subsec:transmission_scheme}
Here, the transmission scheme is mainly focused on physical technologies and interference management at the link level.
At each tier, data is transmitted using a multi-hop strategy from the source to the destination.
Based on the aforementioned $D$-hop maximum routing policy, data is forwarded from one $l$-th tier node to its nearest $(l+1)$-th tier node via multi-hop communication in the upstream transmission stage, and vice versa in the downstream transmission with the $(l+1)$-th tier.
Moreover, we assume that one $(l+1)$-th tier node is capable of utilizing Multi-User MIMO (MU-MIMO) to start $M_{l+1}/M_{l}$ routing paths at the same time slot for upstream and downstream transmission.
For the same-tier transmission, one $l$-th tier node can only forward data to another neighboring node per transmission slot.

All three types of data transmission at the $l$-th tier, under the $D$-hop maximum routing policy, take place between two neighboring cells, or in the same cell when a $l$-th tier source node and its corresponding $(l+1)$-th tier destination node happen to be located in the same cell for upstream transmission or vice versa for downstream transmission.
In this way, time or frequency division can be utilized to reduce excessive interference or strong collisions and satisfy the half-duplex (HD) constraint \cite{FG_ISH_IMH}.
To be specific, supposing that the maximum number of interfering cells at the $l$-th tier is ${\Delta c}_{l}$, then a $(1+{\Delta c}_{l})$-TDMA can be used, where the $l$-th cells alternate in becoming active in every one of $(1+{\Delta c}_{l})$ time slots \cite{kumar}.

Nowadays, MIMO has become a key technology to increase the performance of wireless networks by exploiting the degree-of-freedom (DoF) gain, power gain, and diversity gain \cite{David_Tse_MIMO_Book}.
Notably, diversity gain is aimed at improving the reliability of a transmitted signal by spreading the same signal across two or more uncorrelated communication channels, resulting in a boosted performance of the decoding bit error probability (BER), which has little effect on the increase of channel capacity.
Thus, we consider the following two transmission schemes of MIMO use.

\textbf{Spatial Multiplexing:}
When spatial multiplexing is utilized, multiple independent data streams can be transmitted at the same time.
It has been shown in \cite{los} and \cite{CJ_MIMO_Ad_Hoc} that, utilizing spatial multiplexing for data transmission, a DoF gain of $M$ can be obtained for a single-tier mesh network with nodes each equipped with $M$ antennas.
In the general case, a maximum DoF gain of $N_{\text{min}}=\min \{N_t, N_r \}$ can be obtained, where $N_t$ and $N_r$ are the numbers of antennas equipped at the transmitter and the receiver, respectively \cite{David_Tse_MIMO_Book}.
To reduce the excessive inter-tier interference coming from concurrent transmissions of different tiers, we assume that frequency division is used in this scheme, i.e., orthogonal frequency bands are allocated to data transmission of different tiers.
For data transmission of each tier, time division or frequency division is utilized to reduce intra-tier interference and satisfy the half-duplex constraint.
In summary, using the spatial multiplexing scheme can improve the per-node throughput of wireless mesh networks by increasing the transmission rate of each link, such that the throughput degradation due to link-sharing can be compensated to some extent.

\textbf{Beamforming:}
In this scenario, at each hop, the source node transmits a single stream of data to its destination. 
Under the beamforming scheme, it is possible for each pair of communicating nodes to transmit with thin beams, such that the transmission energy is concentrated in the direction of the receiving node and excessive interference to nearby concurrent receiving nodes can be reduced \cite{los}.
Thus, both intra-tier and inter-tier interference can be regarded as upper bounded by a constant, i.e., $O(1)$.
In this way, frequency reuse among multiple tiers is possible.
It has been shown in \cite{los} that, for a single-tier mesh network with nodes each equipped with $M$ antennas, the power gain which can be obtained from beamforming is equal to $M^2$.
In general, an $N_t \times N_r$ MIMO system can provide a power gain of $N_t N_r$ \cite{David_Tse_MIMO_Book}.
Moreover, by utilizing the power gains provided by beamforming, the transmission range of wireless nodes can be increased such that data transmission of each S-D pair in the mesh network can be finished within fewer hops.
Consequently, the link-sharing issue is alleviated and the required relay nodes can be made less than that of the spatial multiplexing scheme. 
To sum up, using beamforming can not only reduce interference but also alleviate the link-sharing issue by enlarging the transmission range. 

\subsection{Achievable Per-Node Throughput}
\label{subsec:multi-tier_throughput}
This section is mainly focused on deriving the achievable per-node throughput of the spatial multiplexing and beamforming schemes in the multi-tier mesh network under the $D$-hop maximum routing policy, respectively.
The related theorems are stated as follows.

\begin{theorem} (Spatial Multiplexing)
\label{thm:R(n)_SM}
For the proposed multi-tier hierarchical architecture for a mesh network of $n$ data nodes with network parameters scaling according to Table \ref{tab:params_scaling}, the per-node throughput under the $D$-hop maximum routing policy using the spatial multiplexing scheme is
\begin{equation*}
    R^{(SM)}(n) = \min_{1 \le l \le L} \left\{ \Theta \left( l^{\psi + \upsilon - k} \right) \right\}.
\end{equation*}
\end{theorem}

\begin{theorem} (Beamforming)
\label{thm:R(n)_BF}
For the proposed multi-tier hierarchical architecture for a mesh network of $n$ data nodes with network parameters scaling according to Table \ref{tab:params_scaling}, the per-node throughput under the $D$-hop maximum routing policy using the beamforming scheme is
\begin{equation*}
    R^{(BF)}(n) = \min_{1 \le l \le L} \left\{ \Theta \left( \upsilon l^{\psi+ 2 \upsilon / \alpha_l - k} \log l\right) \right\}.
\end{equation*}
\end{theorem}

The detailed processes of deriving $R^{(SM)}(n)$ and $R^{(BF)}(n)$ are demonstrated in the following.

\subsubsection{Spatial Multiplexing}
Based on the aforementioned procedure of deriving the per-node throughput of the multi-tier mesh network, the first thing to do is to determine the number of data flows that pass through each tier.
For simplicity, we consider the data flow of a randomly chosen S-D pair and analyze the probability that the data flow of this randomly chosen S-D pair passes through the $l$-th tier, denoted by $Q_l$, under the $D$-hop maximum routing policy.
The average number of data flows going across the $l$-th tier, denoted by $N_l$, can be then obtained as
\begin{equation}
    \label{equ:N_l_ini}
    N_l = Q_l N_{S-D}.
\end{equation}

Under the $D$-hop maximum routing policy, a data flow will not consume the $l$-th tier's resources if it doesn't go through the $l$-th tier.
The induction method is adopted to derive $Q_l$.
Apparently, $Q_1 = 1$ since the first hop of every data flow is always carried out in the first tier.
Under the $D$-hop maximum routing policy, data will be transmitted to the next higher tier by upstream transmission if the number of hops between the source and destination at the $l$-th tier is larger than $D_l$. 

Let $P_h(h_l=x)$ denote the probability that the hop distance between two communicating nodes at the $l$-th tier is $x$. 
From Fig.~\ref{fig:network}, it can be observed that there are $6x$ nodes that are $x$ hops away from each node at the $l$-th tier without consideration of edge effect \cite{social}.
The edge effect can be ignored when the number of nodes is very large, which holds valid for the lower tiers.
Nevertheless, the node number for a higher tier is a finite integer, which calls the requirement to consider the edge effect when deriving $P_h(h_l=x)$ for large $l$.
Consider the data transmission in the first tier with $n$ data nodes.
Let $b_{i}(n, x)$ denote the number of nodes that are $x$ hops away from node $i$ in a single-tier mesh network with $n$ nodes.
Under the assumption of random and uniform distribution of data flows, i.e., each source node randomly chooses another node as its destination in the $l$-th tier, it can be obtained that
\begin{equation*}
    \label{equ:P_h(D_1=x)_ini}
    P_h(h_l=x) = \frac{1}{n_l} \sum_{i=1}^{n_l} \frac{{b_{i}(n_l,x)}}{n_l-1},
\end{equation*}
which is a constant determined by $x$ and $n$.
The probability that a randomly chosen data flow going across the $l$-th tier can be finished in the $l$ tier under $D$-hop maximum routing policy, denoted by $\xi_l(n_l, D_l)$, is equal to the probability that the hop distance between two communicating nodes is no more than $D_l$, i.e., 
\begin{equation*}
    \xi_l \left(n_l, D_l \right) = \sum_{x=1}^{D_l}P_h(h_l=x) \equiv \xi_l,
\end{equation*}
which is a constant when $n_l$ and $D_l$ are fixed.
It can be seen that $\xi_l$ represents the probability that a data flow requires no more than $D_l$ hops to transmit in the $l$-th tier.
In other words, for data flows going across the $l$-th ($l<L$) tier, the probability that it needs to be transmitted to the $(l+1)$-th tier in upstream transmissions is $(1-\xi_l)$.
Thus, the probability that the data flow of an S-D pair going across the $l$-th tier $Q_l$ can be derived by recurrence.
To be specific, it can be obtained that 
\begin{equation*}
        Q_{l+1} = (1-\xi_l) \cdot Q_l, \quad l = 1, ..., L-1.
\end{equation*}
Since $Q_1=1$, the following expression for $Q_l$ can be derived, 
\begin{equation*} 
    Q_l = \begin{cases}
        1, & l = 1, \\
        \prod_{i=1}^{l-1} \left( 1- \xi_i \right), & l = 2, ..., L-1.
    \end{cases}
\end{equation*}

Let $H_{j,l}$ denote the number of hops for data transmission of a data flow $j$ at the $l$-th tier if it goes across the $l$-th tier. 
Similar to the single-tier case, each node in the multi-tier mesh network resort to the time division method to separate data transfers of multiple data flows it needs to relay.

Under the $D$-hop maximum routing policy, the number of hops in each of the $L$ tiers for all the $n$ data flows will never exceed $D$.
As for the $l$-th tier, there are $N_l = n Q_l$ data flows going across it.
Among them, $N_{l+1} = n Q_{l+1}$ data flows need to be transmitted to the $(l+1)$-th tier through upstream transmissions, leaving $N_{l}^{\prime} = n Q_l^{\prime} = n (Q_{l} - Q_{l+1} )$ data flows transferred using same-tier multi-hop transmission at the $l$-th tier.
As for those $N_{l}^{\prime}$ data flows transferred by the same-tier multi-hop scheme at the $l$-th tier, it can be easily observed that the number of hops between the source and destination node is $\Theta( D_l )$ under the $D$-hop maximum routing policy.

Since those $N_{l+1}$ data flows transmitted to the $(l+1)$-th tier by upstream transmission will still be sent back to the $l$-th tier in downstream transmission from the $(l+1)$-th tier, the number of hops for these data flows at the $l$-th tier can be divided into two parts.
One is the number of hops from the source node at the $l$-th tier to its nearest $(l+1)$-th tier node, the other is the number of hops from the destination node at the $l$-th tier to its nearest $(l+1)$-th tier node.
Recall that one $(l+1)$-th tier node covers the area of $k$ cells that contain the $l$-th tier nodes.
As can be seen, the average number of $l$-th tier cells contained in one $(l+1)$-th tier is
\begin{equation*}
    {\Delta k}_l = \frac{A_{l+1}}{A_{l}} = \left( 1+\frac{1}{l} \right) ^ {k}.
\end{equation*}
Thus, the expected number of hops needed for an $l$-th tier node to access an $(l+1)$-th tier is $\Theta(\sqrt{{\Delta k}_l})$.
The expected value of $H_{j,l}$, under $D$-hop maximum routing policy using spatial multiplexing, can be expressed as
\begin{equation}
    \label{equ:E_H_j,l^SM}
    \mathbb{E}\left[H_{j,l}^{(SM)}\right] = \begin{cases}
        \zeta_{l}^{(u)} \sqrt{{\Delta k}_l}, & \text{for upstream data flows},\\
        \zeta_{l}^{(s)} D_l, & \text{for same-tier data flows},
    \end{cases}
\end{equation}
where $\zeta_{l}^{(u)}$ and $\zeta_{l}^{(s)}$ are two constants with scaling of $O(1)$.

Let $Z_{l,i}$ denote the number of data flows node $i$ at the $l$-th tier needs to help transfer under the $D$-hop maximum routing policy in the multi-tier mesh network. 
Based on the equation that the total number of data flows for the $n_l$ nodes at the $l$-th tier to transfer is equal to the summation of the number of hops at the $l$-th tier, it can be obtained that 
\begin{equation*}
    \sum_{i=1}^{n_l} Z_{l,i} = \sum_{j=1}^{N_l} H_{j,l}.
\end{equation*}
Taking expectations on both sides yields that
\begin{equation}
    \label{equ:EZ_EH}
    \sum_{i=1}^{n_l} \mathbb{E}\left[ Z_{l,i} \right]  = \sum_{j=1}^{N_l} \mathbb{E}\left[ H_{j,l} \right].
\end{equation}
Thus, the number of data flows node $i$ at the $l$-th tier needs to transfer, under the spatial multiplexing scheme, can be obtained as
\begin{equation*}
    \label{equ:E(Z_l,i^SM)}
    \begin{aligned}
        \mathbb{E}\left[ Z_{l,i}^{(SM)} \right] 
        &= \frac{n Q_{l+1} \zeta_{l}^{(u)} \sqrt{{\Delta k}_l} + n (Q_l - Q_{l+1}) \zeta_{l}^{(s)} D_l } {n / l^{k}} \\
        &= l^{k} \left( Q_{l+1} \zeta_{l}^{(u)} \sqrt{{\Delta k}_l}  + (Q_l - Q_{l+1}) \zeta_{l}^{(s)} D_l \right).
    \end{aligned}
\end{equation*}
Since $Q_l$, $Q_{l+1}$, $\zeta_{l}^{(u)}$, and $\zeta_{l}^{(s)}$ are all constants, to minimize the number of data flows for each node to transfer, $D_l$ should be chosen as $D_l = \Theta ( \sqrt{{\Delta k}_l} )$. 
For instance, we choose
\begin{equation}
    \label{equ:D_l}
    D_l = \sqrt{{\Delta k}_l}  = \left( 1+\frac{1}{l} \right)^{k/2}.
\end{equation}
Subscribing \eqref{equ:D_l} back yields
\begin{equation*}
    \mathbb{E}\left[ Z_{l,i}^{(SM)} \right] = \Theta \left( \left( l^2+l \right) ^{k/2} \right) = \Theta \left( l^k \right).
\end{equation*}

As the number of hops for each data flow at the $l$-th tier is upper bounded by a constant under the $D$-hop maximum routing policy, the same-tier multi-hop transmission is finished in a local area, and so are the upstream transmission and downstream transmission.
In this way, an excessive traffic load burden can be rarely observed at each node w.h.p.
Thus, the upper bound on the number of data flows any node at the $l$-th tier needs to transfer $Z_{l}^{(U)}$, under the spatial multiplexing scheme, can be taken as the same order with $\mathbb{E}\left[ Z_{l,i}^{(SM)} \right]$, i.e.,
\begin{equation}
    \label{equ:Z_l_U^SM}
    Z_{l}^{(U, SM)} = \Theta \left( l^{k} \right).
\end{equation}

Since all three types of data transmission of each tier take place in the same cell or between two neighboring cells, the maximum number of interfering neighboring cells for a specific link is upper bounded by a constant, i.e.,
\begin{equation}
    \label{equ:Delta_c_l^SM}
    {\Delta c}_l ^{(SM)} = O \left( 1 \right).
\end{equation}


When orthogonal frequency bands are allocated to different tiers for data transmission, inter-tier interference under the spatial multiplexing scheme is completely canceled.
To be specific, the bandwidth for data transmissions of the $l$-th tier is $W_l$.
Considering the three types of data transmission in the $l$-th tier, the same-tier transmission can be seen as a point-to-point $M_l \times M_l$ MIMO system, while the upstream transmission to the $(l+1)$-th tier, and downstream transmission from the $(l+1)$-th tier can be regarded as the uplink and downlink of an MU-MIMO system with $M_{l+1}/M_{l} = (1+1/l)^{\upsilon}$ users and one base station (BS) \cite{FG_ISH_IMH}, where each user has $M_l$ antennas and the BS has $M_{l+1}$ antennas. 
As can be seen, under the $D$-hop maximum routing policy, data transmission at each tier is similar to that of a single-tier mesh network.
Denote by $\gamma_l^{(L,SM)}$ the lower bound on the SINR of the received signal at the $l$-th tier using spatial multiplexing. 
When time division is used to reduce inter-tier interference of the $l$-th tier, i.e., each node is only active in one of every $(1+{\Delta c}_l)$ time slots.
Consequently, data transmission of each tier is interference-limited and the SINR of received signal scales as $\Omega(1)$ \cite{ish}, i.e.,
\begin{equation*}
    \label{equ:gamma_l^L}
    \gamma_l^{(L,SM)} = \Omega(1).
\end{equation*}
This result still holds when frequency division is used, i.e., evenly dividing the bandwidth of the $l$-th tier $W_l$ into $(1+{\Delta c}_l)$ segments and alternately allocating the segments across the network, which shows a similar effect with the time-division strategy.
Since a DoF gain of $N_{\text{min}}=\min \{ N_t, N_r\}$ can be obtained in an $N_t \times N_r$ MIMO system when spatial multiplexing scheme is utilized \cite{David_Tse_MIMO_Book}, an achievable link rate for the data transmission in the $l$-th tier, denoted by $R_l^{(L,SM)}$, can be obtained as
\begin{equation}
    \label{equ:R_l^SM}
    R_l^{(L,SM)} = \Theta \left( M_l W_l \right) = \Theta \left (M_1 W_1 l^{\psi + \upsilon}. \right).
\end{equation}

Plugging (\ref{equ:Z_l_U^SM}), (\ref{equ:Delta_c_l^SM}), and (\ref{equ:R_l^SM}) into (\ref{equ:R_l(n)_ini}), an achievable end-to-end rate for the $l$-th tier under the frequency division scheme can be obtained as
\begin{equation}
    \label{equ:R_l(n)^SM}
    R_l^{(SM)}(n) = \Theta \left( \frac{M_1 W_1 l^{\psi + \upsilon}}{l^{k}} \right) = \Theta \left( l^{\psi + \upsilon - k} \right).
\end{equation}
Thus, the per-node throughput for the multi-tier mesh network using frequency division among different tiers can be derived by substituting (\ref{equ:R_l(n)^SM}) into (\ref{equ:R(n)_def}), i.e.,
\begin{equation}
    \label{equ:R^SM(n)}
    R^{(SM)}(n) = \min_{1 \le l \le L} \left\{ \Theta \left(l^{\psi + \upsilon - k} \right) \right\},
\end{equation}
which completes the proof on Theorem \ref{thm:R(n)_SM}.

\subsubsection{Beamforming}
In this scenario, it is assumed that all nodes in the multi-tier mesh network utilize beamforming to transmit a single stream of data, where transmission energy is concentrated in a certain direction targeting the receiving node, causing litter interference to other concurrently transmitting nodes.
Utilizing the power gain provided by beamforming, the transmission range of wireless nodes can be increased, as demonstrated in \cite{los}.
To be specific, when each node is equipped with $M$ antennas, a power gain of $M^2$ can be provided.
Then the transmission range can be obtained by solving the equation
\begin{equation*}
    C P M^2 \Tilde{r}_0^{-\alpha} = P^{0},
\end{equation*}
which yields
\begin{equation*}
    \Tilde{r}_0 = \left( \frac{CP}{P^{0}} \right)^{1/\alpha} M^{2/\alpha} = M^{2/\alpha} r_0.
\end{equation*}
As can be seen, the transmission range is increased $M^{2/\alpha}$ times.
In this way, the expected number of hops needed at the $l$-th tier can be reduced to a fraction of that under the spatial multiplexing scheme, i.e.,
\begin{equation}
    \label{equ:E_H_j,l^BF}
        \mathbb{E}\left[H_{j,l}^{(BF)}\right] =
        \frac{\mathbb{E}\left[H_{j,l}^{(SM)}\right]}{M_l^{2/\alpha_l}} = M_l^{-2 / \alpha_l} \mathbb{E}\left[H_{j,l}^{(SM)}\right],
\end{equation}
where $\mathbb{E}\left[H_{j,l}^{(SM)}\right]$ has been derived in (\ref{equ:E_H_j,l^SM}).
Substituting (\ref{equ:E_H_j,l^BF}) into (\ref{equ:EZ_EH}) yields that
\begin{equation*}
    \mathbb{E}\left[H_{j,l}^{(BF)} \right] = \Theta \left(l^{k} M_l^{-2 / \alpha_l}\right) = \Theta \left(l^{k- 2\upsilon / \alpha_l }  \right).
\end{equation*}
Similarly, an upper bound on the number of data flows any node at the $l$-th tier needs to transfer $Z_{l}^{(U)}$, under the beamforming scheme, can be taken as 
\begin{equation}
    \label{equ:Z_l_U^BF}
    Z_{l}^{(U, BF)} = \Theta \left(l^{k-2 \upsilon / \alpha_l }  \right).
\end{equation}
Compared to (\ref{equ:Z_l_U^SM}), it can be seen that
\begin{equation*}
    Z_{l}^{(U, BF)} \le Z_{l}^{(U, SM)}, 
\end{equation*}
since $\alpha_l$ and $\upsilon$ are both positive constants.
Hence, each node needs to help transfer fewer data flows using beamforming, i.e., the link-sharing issue is further alleviated.

Similarly, since data transmissions of each tier are carried out between two neighboring cells or in the same cell, the maximum number of interfering neighboring cells to reduce excessive interference and satisfy the half-duplex constraint, for a specific link under the beamforming scheme, is also upper-bounded by a constant, i.e.,
\begin{equation}
    \label{equ:Delta_c_l^BF}
    {\Delta c}_l ^{(BF)} = O \left( 1 \right).
\end{equation}

Consider a high-SNR environment where noise power is negligible compared to transmit power $P_l$ for $l=1,...,L$.
When beamforming is utilized, it can be regarded that the interference suffered by each receiving node is much too smaller compared to the transmit power, i.e., interference power scales as $O(1)$.
Let $\gamma_l^{(L,BF)}$ denote the lower bound on the SINR of the received signal at the $l$-th tier using the beamforming scheme. 
It can be derived that
\begin{equation*}
    \gamma_l^{(L,BF)} = \Theta (C_l M_l^{2} P_l  d_l^{-\alpha_l}),
\end{equation*}
where $d_l$ represents the expected distance between the transmitter and the receiver of the $l$-th tier, $\alpha_l$ and $C_l$ are path-loss exponent and the characteristic constant determined by frequency, antenna profile, etc. of the $l$-th tier, respectively.
Since an $N_t \times N_r$ MIMO system can provide a power gain of $N_t N_r$ when used for beamforming, a lower bound on link transmission rate, denoted by $R_l^{(L, BF)}$, can be derived under the high-SNR assumption as
\begin{equation}
    \label{R_l^BF}
    \begin{aligned}
        R_l^{(L, BF)} 
        &= \Theta \left( W_l \log \left(C_l M_l^2 P_l d_l^{-\alpha_l}\right) \right) \\
        &\overset{(a)}{=} \Theta \left( W_1 l^{\psi} \log \left( M_1^2 l^{2\upsilon} P_l^{0} \right) \right) \\
        &= \Theta \left( \upsilon l^{\psi} \log  l \right),
    \end{aligned}
\end{equation}
where $(a)$ comes directly from (\ref{equ:P_l_con_ini}).

Plugging (\ref{equ:Z_l_U^BF}), (\ref{equ:Delta_c_l^BF}), and (\ref{R_l^BF}) into (\ref{equ:R_l(n)_ini}), an achievable link rate for the $l$-th tier using beamforming, denoted by $R_l^{(BF)}(n)$, can be obtained as
\begin{equation}
    \label{equ:R_l(n)^BF}
    \begin{aligned}
        R_l^{(BF)}(n) &= \Theta \left( \frac{\upsilon l^{\psi} \log l}{l^{k - 2 \upsilon / \alpha_l }} \right) \\
        &= \Theta \left( \upsilon l^{\psi+ 2 \upsilon/ \alpha_l -k} \log l \right).
    \end{aligned}
\end{equation}
Substituting (\ref{equ:R_l(n)^BF}) into (\ref{equ:R(n)_def}), the per-node throughput for the multi-tier mesh network under the beamforming scheme can be obtained as
\begin{equation}
    \label{equ:R^BF(n)}
    R^{(BF)}(n) = \min_{1 \le l \le L} \left\{ \Theta \left( \upsilon l^{\psi + 2 \upsilon / \alpha_l - k} \log l\right) \right\},
\end{equation}
which completes the proof on Theorem \ref{thm:R(n)_BF}.

\section{Scalability Conditions for A Multi-Tier Mesh Network}
\label{sec:scalability}

In the previous study, we obtained the per-node throughput of a multi-tier mesh network under both spatial multiplexing and beamforming schemes.
The results are subject to the scaling orders of bandwidth allocation, transmit power, and antenna number. 
In this section, we aim to investigate the requirements of such parameters to achieve scalability under different transmission schemes. 
As shown in (\ref{equ:R(n)_def}), the per-node throughput of a multi-tier mesh network $R(n)$ is defined as the minima of the end-to-end rate $R_l(n)$ for $1 \le l \le L$.
To obtain a scalable per-node throughput of $\Theta(1)$, $R_l(n)=\Omega(1)$ should be met for $1 \le l \le L$, i.e., the conditions for a multi-tier mesh network to achieve throughput-scalability are
\begin{equation}
    \label{equ:scalability_conditions}
    R_l(n) = \Omega (1), \quad l=1,2,...,L.
\end{equation}
In the following, the specific scalability conditions for a multi-tier mesh network under spatial multiplexing and beamforming schemes are derived, respectively.

\subsection{Scalability Conditions for Spatial Multiplexing}
The end-to-end rate of the $l$-th tier using spatial multiplexing scheme $R_l^{(SM)}(n)$ has been derived in (\ref{equ:R_l(n)^SM}).
As can be seen, $R_l^{(SM)}(n)$ scales in the order of $l$ to the power of $\psi + \upsilon - k$, which is determined by the scaling relationships of bandwidth, antenna number, and node number.
Apparently, for the first tier with $l=1$, the end-to-end rate is $R_1^{(SM)}(n) = \Theta(M_1 W_1) = \Theta(1)$, which maintains scalable if the allocated bandwidth $W_1$ and antenna number $M_1$ are fixed.
As for upper tiers, substituting 
\begin{equation*}
    R_l^{(SM)}(n)=\Theta(l^{\psi + \upsilon - k})
\end{equation*}
into (\ref{equ:scalability_conditions}), to achieve a scalable throughput using the spatial multiplexing scheme, it should be satisfied that
\begin{equation}
    \label{equ:con_SM}
    \psi + \upsilon \ge k,
\end{equation}
in order to make sure that $R_l^{(SM)}(n) = \Omega (1)$, since in this way, 
\begin{equation*}
    l^{\psi + \upsilon - k} \ge 1, 
\end{equation*}
and hence the per-node throughput of the multi-tier mesh network, obtained by taking the minimum value of $R_l^{(SM)}(n)$ from $l=1$ to $L$, is scalable as $n$ increases.

Notably, (\ref{equ:con_SM}) suggests that to achieve a scalable throughput, higher relay tiers should not become the bottleneck of the network.
They are required to be capable of handling data transfers coming from the lower tiers.
To achieve this, the summation of bandwidth increasing order $\psi$ and antenna number increasing order $\upsilon$, i.e., $\psi + \upsilon$, should be no less than the decreasing order of the node number $k$.
On the other hand, to obtain a scalable throughput, the scaling order for the number of required relay nodes deployed at each tier $k$ should be at least equal to $\psi + \upsilon$.

\subsection{Scalability Conditions for Beamforming}
The end-to-end rate of the $l$-th tier using beamforming scheme $R_l^{(BF)}(n)$ has been obtained in (\ref{equ:R_l(n)^BF}).
As can be seen, $R_l^{(BF)}(n)$ scales in the order of $l$ to the power of $\psi+2 \upsilon/ \alpha_l - k$ times $\upsilon \log l$.
Similarly, for the first tier with $l=1$, the end-to-end rate is $R_1^{(BF)}(n) = \Theta(W_1 \log {M_1}^2 ) = \Theta(1)$, which remains non-decreasing as long as the allocated bandwidth $W_1$ and antenna number $M_1$ are determined.
Since frequency bands can be shared among multiple tiers under the beamforming scheme, a common path loss exponent, say $\alpha$, can be taken for convenience.
In order to make sure that $R_l^{(BF)}(n) = \Omega (1)$ for upper tiers, substituting 
\begin{equation*}
    R_l^{(BF)}(n)=\Theta \left( \upsilon l^{\psi+ 2 \upsilon / \alpha - k} \log l \right)
\end{equation*}
into (\ref{equ:scalability_conditions}), a scalable throughput using the beamforming scheme can be obtained, as $n$ increases, if
\begin{equation}
    \label{equ:con_BF}
    \psi + 2\upsilon / \alpha \ge k.
\end{equation}
In this way, $R_{l}^{(BF)}(n) \ge R_{l-1}^{(BF)}(n)$ for $l=2,...,L$ and therefore $R^{(BF)}(n) = R_{1}^{(BF)}(n) = \Theta(1)$,
which means that a scalable per-node throughput can be obtained.


From (\ref{equ:con_BF}), to achieve a scalable throughput, the summation of bandwidth increasing order $\psi$ and antenna number increasing order $\upsilon$ times $2 / \alpha$, i.e., $\psi + 2 \upsilon / \alpha$, should be no less than the decreasing scaling order of the node number $k$.
On the other hand, to obtain a scalable throughput, the scaling order for the number of required relay nodes deployed at each tier $k$ should be at least equal to $\psi + 2 \upsilon / \alpha$.

\subsection{Clarification on Scalability of Mesh Networks}
\label{subsec:scalability_def}

Assume $k\ge 2$ is satisfied to ensure the convergence of total number of nodes.
From the scalability conditions of \eqref{equ:con_SM} and \eqref{equ:con_BF}, the required bandwidth and antenna numbers should go to infinity as $n \rightarrow \infty$.
Moreover, \eqref{equ:P_L_con} indicate that the transmit power of the highest tier should also increase exponentially with $n$.
In summary, the $\Theta(1)$ throughput for a continuously expanding network can only be obtained with unlimited resources.
However, it is infeasible and impractical for such an endless expansion of network size in a realistic scenario.
The deployment region for a wireless network is always limited in practice.
When the region size of a wireless network reaches a certain value, it is supposed to access the wired network (i.e., the Internet), and thus establish a hybrid networking architecture for communication.
The capacity scaling laws on hybrid networks have already been investigated in \cite{imesh, ish, iadhoc, liu, 2l, social, hybrid}.
Their results indicate that, by utilizing the high throughput provided by wired links, a scalable end-to-end rate between two wireless nodes is achievable.
However, the throughput-scalability of the wireless part in the hybrid network is not fully addressed.
To resolve this issue, we make the following clarification on the scalability of mesh networks in this paper.
Specifically, we are aiming at deploying a throughput-scalable wireless mesh network under the available resources, while making the network size as large as possible.
In other words, a per-node throughput of $\Theta(1)$ is guaranteed for the wireless part of the network deployed in such a finite region without considering the wired nodes.

\subsection{Case Study}

Here we present a case study to investigate the practicality of the proposed multi-tier mesh architecture in the realistic deployment. 
Consider a circular region populated with $10,000$ data nodes.
We found that choosing $k=8$, $\psi=4$, and $\upsilon=4$ can obtain feasible network parameters under the spatial multiplexing scheme, while maintaining the throughput-scalability.
As shown in Table \ref{tab:case_n=10000}, a three-tier mesh network is established with $n_2 = 10^4/2^{8}=39$ and $n_3 = 10^4/3^{8}=2$.
We assume that the path-loss exponents for all three tiers are $3$, and that the antenna gains are $C_1= 3$ dB, $C_2 = 6$ dB, and $C_3 = 9$, respectively.
If each data node has a transmit power of 1 mW, the minimum receiving signal power to achieve a transmission range of 50 m is $P_1^{0}=-78$ dBm, derived by (\ref{equ:P_l_con_ini}), which is a reasonable value in practice.
The network radius is roughly $50\times \sqrt{10000/\pi} = 2.8$ km.
The required transmission range of the second and third tiers can be derived by (\ref{equ:ratio_d}), i.e., $d_2 = 50 \times 2^4 = 800$ m and $800\times 1.4^4 = 4.05$ km.
The required transmit power of nodes in the second and third tiers are 2 W and 13 W, assuming that $P_2^{0} = P_3^{0} = P_1^{0} =-78$ dBm.
We allocate a bandwidth of 10 MHz to the first tier, then the required bandwidth for the second and third tiers are 160 MHz and 810 MHz, respectively, which are reasonable parameters in B5G or 6G networks where high-frequency bands such as mmWave and terahertz can be leveraged.
In addition, supposing each data node is equipped with a single antenna, the required antenna numbers for relay nodes in the second and third tiers are 16 and 81, respectively, which are also achievable in realistic deployment.
All the above results are summarized in Table \ref{tab:case_n=10000}.


\begin{table}[t]
    \centering
    \caption{Network parameters for $n=10,000$ under the spatial multiplexing scheme.}
    \label{tab:case_n=10000}
    \begin{tabular}{c|c|c|c}
        \hline
        Tier index & 1 & 2 & 3 \\
        \hline
        Number of nodes ($n_l$) & 10,000 & 39 & 2 \\
        Antenna gain ($C_l$) & 3 dB & 6 dB & 9 dB \\
        Transmit power ($P_l$) & 1 mW & 2 W & 13 W \\
        Bandwidth ($W_l$) & 10 MHz & 160 MHz & 810 MHz \\
        Antenna number ($M_l$) & 1 & 16 & 81\\
        \hline
    \end{tabular}
    \vspace{-12pt}
\end{table}


\section{Conclusion}
\label{sec:conclusion}
In this paper, interference and link-sharing were identified as two key factors that limit the scalability of mesh networks.
A multi-tier hierarchical mesh network architecture was developed to resolve the link-sharing issue. 
Combined with certain schemes of interference reduction, a scalable per-node throughput of $\Theta(1)$ was proven to be achievable in the multi-tier mesh network, when certain conditions on bandwidth, antenna numbers, and transmit power are satisfied.
The case study carried out also demonstrated the feasibility of the multi-tier mesh networking in the realistic deployment.
However, the results also indicate that, to attain a scalable capacity in a multi-tier mesh network, either the required bandwidth or the number of relay nodes is highly demanding, which requires further exploration on a more efficient architecture and the corresponding networking methods.

\bibliography{reference}

\end{document}